\documentstyle[12pt]{article} 

\begin{document}
 
\newcommand{\beq}{\begin{equation}}
\newcommand{\eeq}{\end{equation}}
\newcommand{\barr}{\begin{eqnarray}}
\newcommand{\earr}{\end{eqnarray}}

\newcommand{\andy}[1]{ }

\author{{\bf F. Goy}  \\
\\     
Dipartimento di Fisica\\
Universit\^^ {a} di Bari \\
Via G. Amendola 173 \\
I-70126  Bari, Italy \\ 
E-mail: goy@axpba1.ba.infn.it
\\
\\
and
\\
\\
{\bf F. Selleri}\\
\\
Dipartimento di Fisica\\
Universit\^^ {a} di Bari \\
INFN - Sezione di Bari\\ 
Via G. Amendola 173 \\
I-70126  Bari, Italy \\ 
}
        
\title{{\bf Time on a Rotating Platform }}
 
\date{February 25, 1997}
 
\maketitle
 
\begin{abstract}
Traditional clock synchronisation on a rotating platform is shown to be 
incompatible with the experimentally established transformation of time. The 
latter transformation leads directly to solve this problem through noninvariant
one-way speed of light. The conventionality of some features of relativity 
theory allows full compatibility with existing experimental evidence. 
\\
\\
{\bf Keywords}: Relativity (Special and General), Synchronisation, Sagnac 
effect.
\end{abstract}
 
 
\setcounter{equation}{0}

\section{Muon Storage-Ring experiment}

We start by recalling a well known experiment in which the relativistic 
approach works perfectly, and take from it two lessons concerning the 
transformation of time between the laboratory and a rotating platform.

Lifetimes of positive and negative muons were measured in the CERN Storage-Ring 
experiment \cite{bail:77a} for muon speed $0.9994 c$, corresponding to a 
$\gamma$ factor of $29.33$. Muons circulated in a $14$ m diameter ring, with an 
acceleration of $10^{18}$g. Excellent agreement was found with the relativistic 
formula
\beq
\tau_{0}=\frac{\tau_{rest}}{\sqrt{1-\beta^{2}}} 
\label{eq:eq1}
\eeq
where $\tau_{0}$ is the observed muon lifetime, $\tau_{rest}$ is the lifetime
of muons at rest, and $\beta=v/c$, $v$ being the laboratory speed of the muon
on its circular orbit.

Consider an ideal platform rotating with the same angular velocity as the muon 
in the e.m. field (with respect to such a platform the muon is at rest). 
Consider also four different observers:
\begin{enumerate}
\item $O_{L}$ is the observer in the laboratory reference frame $S_{L}$, 
assumed to be an inertial frame. Thus $O_{L}$ could be the CERN storage ring
experimenter.
\item $O_{a}$ is the accelerated but localized observer who lives on the rim 
of the platform $S_{a}$, very near the muon which looks constantly at rest to 
him; $O_{a}$ has a local knowledge of the platform and of its physical 
properties extending only to the immediate surrounding of his position;
\item  $\widetilde{O_{a}}$ is a second accelerated observer. He knows 
everything 
about the platform (the accelerated frame $S_{a}$) through which he can freely 
move;
\item $O_{T}$ is an observer living in an inertial frame $S_{T}$ in which at a 
certain time $O_{a}$ and the muon are instantaneously at rest. $S_{T}$ will be
called the ``tangeant'' inertial frame.
\end{enumerate}

We give now the description of the muon lifetime from the point of view of 
these four observers.

\begin{description}

\item[D1] According to $O_{L}$ the muon lifetime $\tau_{0}$ is greatly enhanced 
with respect to that ($\tau_{rest}$) of muons at rest in $S_{L}$. His 
observations are expressed by Eq. (\ref{eq:eq1}).
\item[D2] According to $O_{a}$, who knows only the time marked by his local 
clock, the muon lifetime is $\tau_{rest}$. Of course $O_{a}$ is under the 
action of a large acceleration ($10^{18}$g), which he detects as a radial 
gravitational field, but nevertheless his lifetime measurement give just 
$\tau_{rest}$, as for muons at rest in $S_{L}$ observed by $O_{L}$.
\item[D3] According to the accelerated observer $\widetilde{O_{a}}$ the 
clocks on the platform have 
a pace dependent on their position, the fastest going one being
that in the center;
in agreement with the equivalence principle he attributes this phenomenon to 
the presence of a position-dependent radial gravitational field of cosmic 
origin. He can check that the lifetime of muons near the rim of the platform 
is either $\tau_{0}$ or $\tau_{rest}$ depending on the clock chosen (in the 
center or near the rim respectively) for measuring it. Therefore he explains 
the 
value $\tau_{rest}$ found by $O_{a}$ as a consequence of the cosmic 
gravitational field delaying in the same way muon decay and the clock used by
$O_{a}$ for lifetime measurements.
\item[D4] According to the observer $O_{T}$ belonging to the tangent inertial 
frame $S_{T}$ the lifetime is $\tau_{rest}$ (measured of course for muons at 
rest in his frame).
\end{description}

The first lesson to be learnt from the previous conclusions concerns the 
transformation of time given by Eq. (\ref{eq:eq1}): {\em The laboratory
time interval} $\Delta t_{0}$ {\em between two events taking place in a fixed 
position 
on the rotating disk} (muon injection and decay, in the previous example) {\em 
is
seen dilated by the usual  relativistic factor compared with the corresponding 
time interval} $\Delta t$ {\em measured by the accelerated observer} $O_{a}$:
\beq
\Delta t_{0}=\frac{\Delta t}{\sqrt{1-\beta^{2}}} 
\label{eq:eq2}
\eeq
 
The second lesson is that the observers $O_{a}$ {\em and} $O_{T}$ {\em agree 
on the 
laws of nature, for example on the decay rate of muons at rest, even though}
$O_{a}$ {\em feels the presence of a radial gravitational field (of cosmic 
origin) 
while} $O_{T}$ {\em does not}. Of course, this conclusion is not new. For 
example, Einstein \cite{eins:11a}, M\mbox{\o}ller \cite{moll:72a}, and Vigier 
\cite{vigi:95a} assumed that the acceleration of a clock $C_{a}$ relative to an 
inertial system has no influence on the rate of $C_{a}$, and that the increase 
in the proper time of $C_{a}$ at any time is the same as that of  the standard 
clocks of the inertial system in which $C_{a}$ is momentarily at rest. Of 
course the situation is different for an observer considering acceleration due 
to a gravitational field, as shown above. The identical conclusions of $O_{a}$ 
and $O_{T}$ imply that the speed of light found locally in the accelerated 
system should be the same as that observed in the ``tangeant'' inertial frame.
But in special relativity the latter speed is always $c$ and we are so brought 
to conclude that the speed of light relative to an accelerated system should 
be also $c$. However this conclusion gives rise to endless trouble.  

\section{The traditional clock synchronisation procedure}

If a disk is rotating with constant angular velocity with respect to an 
inertial frame, one 
can obtain the metric on the disk as follows: in the inertial system the 
invariant squared space-time distance $ds^{2}$ in cartesian coordinates is
\beq
ds^{2}= c^{2}dt_{0}^{2}-dx_{0}^{2}-dy_{0}^{2}-dz_{0}^{2}
\label{eq:eq3}
\eeq
In general relativity one is free to adopt any set of coordinates useful for 
solving a given problem, independently of their physical meaning. In the case 
of the rotating disk it is simpler to use the coordinates in the right-hand
side of the following transformations, as done for example by Langevin 
\cite{lang:21a} and by some texbooks \cite{lali:59a}
\barr
t_{0}&=&t \nonumber\\
x_{0}&=& r \cos(\varphi+\omega t) \nonumber\\
y_{0}&=& r \sin(\varphi+\omega t) \nonumber\\
z_{0}&=&z
\label{eq:eq4}
\earr
The variables $t,\; r,\; \varphi,\;z$ give {\em a possible} 
(although not the best) 
description of physical events for an observer at rest on the disk. In 
(\ref{eq:eq4}) simplicity is attained at the price of provisionally neglecting 
time dilation and lenght contraction. By substituing (\ref{eq:eq4}) in 
(\ref{eq:eq3}) one can easily obtain:
\beq
ds^{2}=(1-\omega^{2}r^{2}/c^{2})(c\;dt)^{2}-
\frac{2\omega r^{2}}{c}d\varphi(c\;dt)
-dz^{2}-dr^{2}-r^{2}d\varphi^{2}
\label{eq:eq5}
\eeq
Eq. (\ref{eq:eq5}) defines a metric $g_{ij}$ which is stationary, but not 
static. If $x^{0}=ct$, $x^{1}=r$, $x^{2}=\varphi$, $x^{3}=z$, its element are
\barr
g_{00}&=& 1-\omega^{2}r^{2}/c^{2}\nonumber\\
g_{11}&=&g_{33}=-1 \nonumber \\
g_{20}&=&g_{02}= -r^{2}\omega/c \nonumber\\
g_{22}&=&-r^{2}
\label{eq:eq6}
\earr
all other elements being zero. Note that the space-time described by 
(\ref{eq:eq6}) is flat because $R_{ijkl}(t',x',y',z')=0 \Rightarrow
R_{ijkl}(t,r,\varphi,z)=0\;[i,j,k,l=0,1,2,3]$, where $R_{ijkl}$ is the 
Riemann 
tensor. For the same reason of covariance the metric defined in (\ref{eq:eq6}) 
is necessarily a solution of the Einstein equations in empty space $R_{ij}=0
\;[i,j=0,1,2,3]$, where $R_{ij}$ is the Ricci tensor.

The proper time differential $d\tau$ of a clock located in a fixed point of the 
disk of space coordinates $(r,\varphi,z)$ is obtained by equalling all space 
differentials to $0$ and taking into account that $d\tau=ds/c$, so that:
\beq
d\tau=\sqrt{g_{00}}\;dt=\sqrt{1-\omega^{2}r^{2}/c^{2}}\; dt
\label{eq:eq7}
\eeq

The length element $dl$ between a point $A$ of the three dimensional space with 
coordinates $x^{\alpha}+dx^{\alpha}\;\;[\alpha=1,2,3]$ and an infinitesimally 
near point $B$ of coordinates $x^{\alpha}$ is found by sending a light signal
from $B$ to $A$ and back, and assuming that the {\em two-way} velocity of 
light is $c$ in all directions, a safe procedure from our point of view. The 
proper time in $B$ needed for this operation (multiplied by $c$) is by 
definition twice the length $dl$ between $A$ and $B$. It is found that:
\beq
dl^{2}=(
-g_{\alpha\beta}+\frac{g_{0\alpha}g_{0\beta}}{g_{00}})dx^{\alpha}dx^{\beta}=
dr^{2}+dz^{2} + \frac{r^{2}d\varphi^{2}}{1-\omega^{2}r^{2}/c^{2}}
\label{eq:eq8}
\eeq
Note the $r$-dependent coefficient of $r^{2}d\varphi^{2}$: {\em space} is not 
flat.

In relativity an observer on a rotating platform $O_{a}$ must synchronise 
clocks placed in different points {\em by assuming that the one-way velocity 
of light is} $c$ {\em in all directions of his noninertial frame}. This 
assumption is in agreement with the ``second lesson'' taken from the muon 
experiment, but becomes the source of a big problem which makes experts 
conclude that ``the rotating platform in relativity is a mystery''. By applying 
it, the ``time'' $t_{B}$ in $B$ is called synchronous with the ``time'' $t_{A}$
in an infinitesimally near point $A$ when
\beq
t_{B}=t_{A} + \Delta t= t_{A} -\frac{1}{c}\frac{g_{0\alpha}dx^{\alpha}}{g_{00}}
=t_{A}+\frac{\omega r^{2}d\varphi}{c^{2}(1-\omega^{2}r^{2}/c^{2})}
\label{eq:eq9}
\eeq
This definition is equivalent to the standard synchronisation procedure of SRT 
and  is obtained by assuming that the point $B$ receives a signal from $A$ 
exactly at midtime between the times of light departure from $A$ and return 
in $A$. The assumption is that the one-way velocity of light is $c$.

Notice that the points $A$ and $B$ chosen for defining the synchronisation
(\ref{eq:eq9}) are totally arbitrary: in general they are not at the same 
distance $r$ from the centre. The consequence is that $\Delta t$ is not a total 
differential. In fact for all functions $f(r,\varphi)$ one has
\beq
\Delta t=\frac{\omega r^{2} d\varphi}{c^{2}(1-\omega^{2}r^{2}/c^{2})} \neq 
df(r,\varphi)
\label{eq:eq10}
\eeq
The proof is very simple. In fact, $\Delta t$ is proportional to $d\varphi$ and 
does not contain $dr$. The coefficient of $d\varphi$ however 
does contain $r$. The 
two statements are incompatible for total differential of any $f(r,\varphi)$. A 
diffeomorphism of the type:
\[
T\;:\;\;\left\{ 
\begin{array}{ccl}
t & \mapsto & \tilde{t}=f_{1}(r,\varphi)t+f_{2}(r,\varphi)\\
r & \mapsto & \tilde{r}= f_{3}(r,\varphi)\\
\varphi & \mapsto & \tilde{\varphi}=f_{4}(r,\varphi)
\end{array}
\right.
\]
is unable to transform $\Delta t$ in a total differential (proof available by 
the authors upon request). [Physical meaning of $T$: the spatial part is chosen 
time independant, so that we remain on the disk; time undergoes the most 
general linear transformation; z play no role]. The latter result agrees with 
Landau and Lifschitz \cite{lali:59a} who stated that the inequality 
(\ref{eq:eq10}) is not dependant on the choice of transformations, but is of 
general validity since
\beq
\Delta t = -\frac{1}{c}\frac{g_{0\alpha}dx^{\alpha}}{g_{00}}
\label{eq:eq11}
\eeq
can be a total differential only in very special and physically uninteresting
cases. M\mbox{\o}ller \cite{moll:72a} and Landau and Lifschitz \cite{lali:59a}
say that it is possible to define the standard synchronisation along a non 
closed curve, but that it is impossible along a closed curve when the metric is 
not static. In fact, given (\ref{eq:eq10}), this synchronisation procedure is 
path dependant, so that one will generally not obtain the same result when 
synchronising a clock $B$ with a clock $A$ using two different paths. This 
means also that if a clock $B$ is synchronised with $A$ and a clock $C$ is 
synchronised with $B$, $C$ will generally not be synchronised with $A$.
This matter was investigated by Anandan \cite{anan:81a} who admitted the 
existence of a ``time lag'' in synchronising clocks around the circle and found 
for it a rather abstract interpretation, and by Ashtekar and Magnon 
\cite{asma:75a} who limited themselves to a formal approach.

The existence of a synchronisation problem is physically strange because if 
the whole disk is initially at rest in the laboratory (inertial) frame $S_{L}$,
with clocks near its rim synchronised with the regular procedure used for all 
clocks of $S_{L}$, then when the disk moves, accelerates, and attains a 
constant angular velocity, the clocks must slow their rates but cannot 
desynchronise for symmetry reasons, since they have at all times the same 
speed. From such a point of view it is difficult to see why there should be any 
difficulty in defining time on the rotating platform. The necessity to 
distinguish sharply between questions of clock phase (distant simultaneity) 
from those of clock rate was stressed by Phipps \cite{phip:95a} with whom we 
fully agree on this point.

\section{Noninvariant speed of light}

The laboratory is assumed to be an inertial frame in which clocks have been 
synchronised with the standard relativistic method.

We consider only clocks on the uniformly rotating platform having radius $R$ 
and angular velocity $\omega$ that are near its rim. We assume them to be 
synchronised as follows: When the clocks of the laboratory show the time 
$t_{0}=0$ then also the clocks on the platform are all set at the time $t=0$.
By symmetry reasons the clocks on the platform will share the following 
property during the uniform rotation: any observer at rest in the laboratory 
near the rim of the platform whose clock marks the time $t_{0}$ will see the 
clock on the platform passing by in that very moment marking the time
\beq
t=t_{0}\sqrt{1-\beta^{2}}
\label{eq:eq12}
\eeq
with $\beta=\omega R/c$.

Near the rim of the platform besides clocks there are also: (i) A light
source $\Sigma$ placed in a fixed position; near $\Sigma$ there is a clock 
$C_{\Sigma}$; (ii) A backward reflecting mirror $M$ placed in diametrically 
opposite position  with respect to $\Sigma$; near $M$ there is a clock $C_{M}$.
At time $t_{1}$ of $C_{\Sigma}$, $\Sigma$ emits a flash of light that propagates
circularly and (we assume) in the direction of rotation of the disk with 
respect to the laboratory, until it arrives at $M$ at time $t_{2}$ of $C_{M}$.
The flash is reflected back, propagates circularly in the opposite direction, 
arrives back at $\Sigma$ at time $t_{3}$ of $C_{\Sigma}$

In the theory of relativity it is {\em assumed} that the one-way velocity of 
light has the same value from $\Sigma$ to $M$ as from $M$ to $\Sigma$, so that
$t_{3}-t_{2}=t_{2}-t_{1}$, whence the $C_{M}$ time $t_{2}$, can be written in 
terms of the two $C_{\Sigma}$ times $t_{1}$ and $t_{3}$ as follows:
\beq
t_{2}=t_{1}+\frac{1}{2}(t_{3}-t_{1})
\label{eq:eq13}
\eeq
Reichenbach commented \cite{reic:58a}: ``{\em This definition is essential 
for the 
special theory of relativity, but it is not epistemologically necessary. If we 
were to follow an arbitrary rule restricted only to the form}
\beq
t_{2}=t_{1}+\varepsilon(t_{3}-t_{1})\;,\;\;\; 0<\varepsilon<1
\label{eq:eq14}
\eeq
{\em it would likewise be adequate and could not be called false. If the 
special theory of relativity prefers the first definition, i.e., sets} 
$\varepsilon$ {\em equal to} $1/2$, {\em it does so on the ground that this 
definition leads to simpler relations''.} On the possibility to choose freely
$\varepsilon$ according to (\ref{eq:eq14}) agreed, among others, Gr\"{u}nbaum 
\cite{grun:73a}, Jammer \cite{jamm:79a}, Mansouri and Sexl \cite{mase:77a}, 
Sj\"{o}din \cite{sjod:79a}, Cavalleri \cite{cava:89a}, Ungar \cite{unga:91a}.

Clearly, different values of $\varepsilon$ correspond to different values of 
the one-way speed of light. In fact, one can write
\beq
t_{2}-t_{1}=\frac{L}{2\tilde{c}(0)}\;\;\; and \;\;\;
t_{3}-t_{2}=\frac{L}{2\tilde{c}(\pi)}
\label{eq:eq15}
\eeq
where $L/2$ is the $\Sigma - M$ distance, $\tilde{c}(0)$ is the one-way 
velocity of light from $\Sigma$ to $M$ and $\tilde{c}(\pi)$ is the one-way 
velocity from $M$ to $\Sigma$. By adding the previous relations one gets
\beq
t_{3}-t_{1} = \frac{L}{2\tilde{c}(0)} +\frac{L}{2\tilde{c}(\pi)}=\frac{L}{c}
\label{eq:eq16}
\eeq
the last step being necessary, because the {\em two-way} velocity of light
has been measured with great precision and always found to be $c$.
From (\ref{eq:eq14}), (\ref{eq:eq15}) and (\ref{eq:eq16}) one easily gets
\beq
\varepsilon=\frac{t_{2}-t_{1}}{t_{3}-t_{1}}=\frac{c}{2\tilde{c}(0)}
\label{eq:eq17}
\eeq
Therefore freedom of choice of $\varepsilon$ means freedom of 
choice of the one-way velocity of light! We believe that it is necessary to 
exploit the free choice of the one-way speed of light, which has never been 
measured, given that the standard assumption $\tilde{c}(0)=\tilde{c}(\pi)=c$ 
leads to contradictions as we saw, and as will become even clearer by the 
following considerations.

The description of the light circulating along the rim of the disk given by the 
laboratory observers will be the following: At time $t_{01}$ the source emits
a light flash that propagates circularly and arrives at $M$ at time $t_{02}$,
is reflected back, propagates circularly, arrives back at $\Sigma$ at time
$t_{03}$. These laboratory times are related to the corresponding platform
times by
\beq
t_{0i}=\frac{t_{i}}{\sqrt{1-\beta^{2}}}\;\;\;\;\; (i=1,2,3)
\label{eq:eq18}
\eeq
as a consequence of (\ref{eq:eq12}).

If $L_{0}$ is the disk circumference length measured in the laboratory, light 
propagating in the rotational direction of the disk must cover a  
distance larger than $L_{0}/2$ by a quantity $x=\omega R(t_{02}-t_{01})$
equalling the shift of $M$ during the time $t_{02}-t_{01}$ taken by light to 
reach $M$. Therefore
\beq
\frac{L_{0}}{2}+x=c(t_{02}-t_{01})\;\;\; ;\;\;\;
x=\omega R(t_{02}-t_{01})
\label{eq:eq19}
\eeq
From these equations it is easy to get:
\beq
t_{02}-t_{01}=\frac{L_{0}}{2c(1-\beta)}
\label{eq:eq20}
\eeq
After reflection light propagates in the direction opposite to that of 
rotation and must now cover a distance smaller than the disk semicircumference 
length $L_{0}/2$ by a quantity $y=\omega R(t_{03}-t_{02})$ equalling the shift 
of $\Sigma$ during the time $t_{03}-t_{02}$ taken by light to reach $\Sigma$.
Therefore
\beq
\frac{L_{0}}{2}-y=c(t_{03}-t_{02})\;\;\; ;\;\;\;
y=\omega R(t_{03}-t_{02})
\label{eq:eq21}
\eeq
One now gets
\beq
t_{03}-t_{02}=\frac{L_{0}}{2c(1+\beta)}
\label{eq:eq22}
\eeq
Summing together (\ref{eq:eq20}) and (\ref{eq:eq22}) it follows
\beq
t_{03}-t_{01}=\frac{L_{0}}{c}\frac{1}{1-\beta^{2}}
\label{eq:eq23}
\eeq
We show next that these relations fix the synchronisation on the disk.
In fact (\ref{eq:eq18}) applied to (\ref{eq:eq20}) and (\ref{eq:eq23}) gives
\beq
t_{2}-t_{1}=\frac{L_{0}\sqrt{1-\beta^{2}}}{2c(1-\beta)}\;\;\; ;
\;\;\; t_{3}-t_{1} = \frac{L_{0}}{c\sqrt{1-\beta^{2}}}
\label{eq:eq24}
\eeq
so that
\beq
\varepsilon=\frac{t_{2}-t_{1}}{t_{3}-t_{1}}=\frac{1+\beta}{2}
\label{eq:eq25}
\eeq
Comparing with (\ref{eq:eq17}) we get
\beq
\tilde{c}(0) = \frac{c}{1+\beta}
\label{eq:eq26}
\eeq
An analogous reasoning made for light emitted by $\Sigma$ in the direction 
opposite to disk rotation leads to
\beq
\tilde{c}(\pi) = \frac{c}{1-\beta}
\label{eq:eq27}
\eeq
Eq.s (\ref{eq:eq26})-(\ref{eq:eq27}) give the one-way speed of light on the 
platform. They are particular cases of the formula 
$\tilde{c}(\theta)= c/(1+\beta \cos \theta)$ discussed at length in Ref.
\cite{sell:95a} and shown to be compatible with experimental evidence at the 
special relativistic level (no accelerations).

\section{The Sagnac effect} 

The reasoning of the previous section was made under the assumption that the 
platform clocks near the rim (the only ones we considered) were all 
synchronised with the laboratory clocks at the same (laboratory) time. This 
procedure clearly amounts also to a synchronisation of the platform clocks with 
respect to one another. It might lead to the incorrect idea that the obtained 
one-way velocity of light different from $c$ is only a consequence of the 
chosen synchronisation. In order to dispel this impression we repeat here the 
reasoning by using only the clock $C_{\Sigma}$ on the platform.

Two light flashes leave $\Sigma$ at time $t_{1}$. The first one propagates 
circularly in the sense opposite to the platform rotation and comes back to 
$\Sigma$ after a $2\pi$ rotation at time $t_{2}$. The second one propagates 
circularly in the same rotational sense of the platform and comes back to 
$\Sigma$ after a $2\pi$ rotation at time $t_{3}$. Quite generally we can write
\beq
t_{2}-t_{1}=\frac{L}{\tilde{c}(\pi)} \;\;\; ; \;\;\;
t_{3}-t_{1}=\frac{L}{\tilde{c}(0)}
\label{eq:eq28}
\eeq
It follows
\beq
t_{3}-t_{2} = L \left[\frac{1}{\tilde{c}(0)}-\frac{1}{\tilde{c}(\pi)}\right]
\label{eq:eq29}
\eeq

Describing the experiment from the point of view of the laboratory observer one 
must give a treatement strictly analogous to that of the previous section. It 
results in 
\beq
t_{03}-t_{01}=\frac{L_{0}}{c(1-\beta)}\;\;\; ;
\;\;\;
t_{02}-t_{01}=\frac{L_{0}}{c(1+\beta)}
\label{eq:eq30}
\eeq
The time delay between the two arrivals back in $\Sigma$  is therefore observed
in the laboratory to be
\beq
t_{03}-t_{02}=\frac{2L_{0}\beta}{c(1-\beta^{2})}
\label{eq:eq31}
\eeq
which is the standard formula for the Sagnac effect. Noticing that 
(\ref{eq:eq31}) is the time difference between two events taking place in the 
same point $\Sigma$ on the disk we can apply what we called the first lesson 
from the muon experiment [Eq. (\ref{eq:eq2})] to the laboratory time interval 
$t_{03}-t_{02}$ and also use (\ref{eq:eq29}) to get
\beq
\frac{1}{\tilde{c}(0)}-\frac{1}{\tilde{c}(\pi)}=\frac{t_{3}-t_{2}}{L}=
2\frac{L_{0}}{L}\frac{\beta}{c\sqrt{1-\beta^{2}}}
\label{eq:eq32}
\eeq
Of course one has $L_{0}=L \sqrt{1-\beta^{2}}\;$: the rotating disk 
circumference 
length appears contracted in the laboratory. Therefore
\beq
\frac{1}{\tilde{c}(0)}-\frac{1}{\tilde{c}(\pi)}=\frac{2\beta}{c}
\label{eq:eq33}
\eeq
It is enough to add to (\ref{eq:eq33}) the condition that the two-way velocity 
of light is $c$:
\beq
\frac{L}{\tilde{c}(0)}+\frac{L}{\tilde{c}(\pi)}=\frac{2 L}{c}
\label{eq:eq34}
\eeq
to arrive again at the results (\ref{eq:eq26}) and (\ref{eq:eq27}). There is 
absolutely no way of obtaining the relativistic condition
 $\tilde{c}(0)=\tilde{c}(\pi)=c$. By accepting (\ref{eq:eq26})-(\ref{eq:eq27})
 we find instead a perfectly rational description of the Sagnac effect on the 
rotating platform and overcome the longstanding ``mystery'' of the rotating 
platorm.

\section{Conclusion}

The Sagnac effect \cite{sagn:13a} is essentially the observation of a phase 
shift between two coherent beams travelling on opposite paths in an 
interferometer placed on a rotating disk. Nowdays the Sagnac effect is observed 
with light (in ring lasers and in fiber optics interferometers \cite{eudd:85a})
and in interferometers built for electrons \cite{hani:93a}, neutrons 
\cite{coov:75a}, atoms \cite{stco:94a} and superconducting Cooper pairs 
\cite{zime:65a}. The phase shift in the interferometers is a consequence of the 
time delay between the arrivals of two beams, so a Sagnac effect is also 
measured directly with atomic clocks timing light beams sent around the earth 
via satellites \cite{alla:85a}. In the typical experiment for the study of the 
effect a monochromatic light source placed on the disk emits two coherent beams 
of light in opposite directions along the disk circumference until they reunite 
in a small region and interfere, after a $2\pi$ propagation. The positioning 
of the interference figure depends on the disk rotational velocity. Textbooks 
deduce the Sagnac formula in the laboratory (essentially our Eq. 
(\ref{eq:eq31}) above), but say nothing about the description of the phenomenon 
on the rotating platform. Exception to this trend are Langevin \cite{lang:21a},
Anandan \cite{anan:81a}, Dieks and Nienhuis \cite{dini:90a}, and Post 
\cite{post:67a}, but dissatisfaction remains widespread, because none of these 
treatments is free of ambiguities. For example Langevin's approach leads to 
all the 
difficulties we discussed in the second section [however in his 1937 paper he 
recognized the possibility of a nonstandard velocity of light on the rotating 
platform and gave formulae which agree to first order with our results
(\ref{eq:eq26})-(\ref{eq:eq27})]. As a second example, Post's relativistic 
formula is not generally valid, but limited to the arbitrary case where the 
origin of the ``tangent'' inertial frame coincides with the centre of the 
rotating disk.

It is well known, especially after the work of Reichenbach \cite{reic:58a} and
Mansouri and Sexl 
\cite{mase:77a}, that clock synchronisation is a purely conventional procedure 
when only inertial frames are involved. In other words one is free to choose
either the standard synchronisation, or a nonstandard one leading 
to a noninvariant velocity of light. Either choice will allow full 
agreement with 
experimental facts. However we have shown that the conventionality of the 
synchronisation procedure is not preserved in accelerated systems, and that a 
theory free of logical contradictions {\em must} choose a one-way velocity of 
light which is nonstandard when measured in the accelerated frame. By the way, 
this is exactly what is already done in practice by physicists synchronising 
clocks around the earth by means of light signals. `` Thus one discards 
Einstein synchronisation in the rotating frame'' said Ashby in the opening talk 
of the 1993 International Frequency Control Symposium \cite{ashb:93a}.

\section{Acknowledgement}

One of us (F.G.) would like to thank the Physics Department of Bari University 
for the temporary hospitality extended to him during the preparation of the 
present paper.


\end{document}